\documentclass[12pt]{article}
\begin{document}
\begin{center}
\large{\bf {Properties, Generations and Masses}}
\vspace{0.2in}

\small{
R Delbourgo\footnote{Bob.Delbourgo@utas.edu.au}

\em{
School of Mathematics and Physics, University of Tasmania,
         Locked Bag 37 GPO Hobart, AUSTRALIA 7001}}
\end{center}
         
\begin{abstract}
Schemes based on anticommuting scalar coordinates, corresponding to properties,
lead to generations of particles very naturally. In contrast to the standard 
model, where masses arise through independent Yukawa couplings to a single
Higgs isodoublet, property models produce Higgs fields in various multiplet
denominations, but contain a single Yukawa coupling. A renormalizable superHiggs 
potential of quartic order then produces very strongly constrained masses
for generations of fields, which depend on just three constants. By allowing for a
small parameter, which is meant to characterise the quantum loop effects for the
effective potential, one can obtain nonzero masses for the engendered masses.
We illustrate the phenomenon for two and three complex coordinates; the more 
realistic case of five complex coordinates is not yet fully treatable because of its 
complexity.
\end{abstract}
\vspace{0.3in}


\section{Masses for generations} %Section I
Two of the most mysterious problems in particle physics are the occurrence of 
generations (like the electron, muon and tauon) and the manner by which their
masses are produced. These problems persist in the standard model where there is
no need for generations, or even a limit on them, and where the arbitrariness
of the several Yukawa couplings of source fields to the (usually single) Higgs 
boson is not constrained. Flavour mixings between families in mass eigenstates 
are yet another source of puzzlement, whether for neutrinos or down-quarks. 
It is therefore not surprising that much research has gone into seeking simple 
models which reduce these ambiguities. Supersymmetry, based on spinorial 
coordinates, is one popular choice but this comes at the expense of at least 
doubling the number of particles and relegating the unseen superpartners into 
unexplored, higher mass regions; other lines of research include grand unified groups, 
non-commuting spacetime coordinates, superstrings, branes,etc. Given that none of these
scheme has received any {\em direct} experimental endorsement, it is not unreasonable 
to look for other avenues which may shed light on these conundrums.

Over the last few years \cite{DTZ88,DW90,DJW91} we have advocated using 
anticommuting Lorentz scalar
coordinates ($\zeta$) as signposts of particle properties and extending 
space-time coordinates ($x$) with these, so that fields become functions both
of location and property: $X = (x,\zeta)$. An expansion of a superfield in $x$ 
and $\zeta$ designates where the particle is and what it is. [The construction must 
be done carefully \cite{D00,D01,D02} so as not to conflict with the spin-statistics theorem.]
Having just a few independent $\zeta$ then leads naturally to a number of
particle generations, without having to invoke some external family group.
Previous investigations have shown that such a scheme based on 5 complex $\zeta$
can accommodate the known fundamental particle spectrum and its repetitions, 
but they have not yet shown how the masses of the multiplets arise. That is the
purpose of the present paper. We will demonstrate that a renormalizable quartic
superHiggs potential having only three constants can lead to masses of the
sources and scalars which are strongly tied to one another at the semiclassical
level. This makes it a good scheme, in that it lends itself to falsification.
The more significant point about the present work is that expansion of the 
superHiggs field $\Phi(X)$ into powers of $\zeta$ gives rise to several Higgs
multiplets, a few of which can have zero-charge expectation values; but there is
only a {\em single} Yukawa coupling to the superfield source field $\Psi(X)$,
so all the masses are fixed by just a few parameters. Of course it is still
unclear how quantum loop corrections can alter the results, but as one is dealing
with a renormalizable model those effects are in principle controllable or at 
least can be related to one another. We shall introduce a small parameter scale $\Delta$
to account in a crude way for the loop terms in the effective action; otherwise one 
simply ends up with zero masses, which is unsatisfactory.

We start by studying a purely leptonic model having but two independent (complex) 
properties $\zeta$. This scheme has only one lepton generation, upon imposing 
self-duality constraints on the $\Phi, \Psi$ expansions in powers of $\zeta$. With 
such a model the Higgs fields are somewhat unrealistic, being weak isocalar and 
isovector; nonetheless the two fermions can in principle be separated in mass scale 
and they are strongly tied to the two Higgs field masses. In the next section we investigate
a more realistic scenario by attaching a third (colourless) baryonic coordinate, 
whereupon the superHiggs expansion yields a weak isodoublet Higgs field plus
three isocalars and one isovector. This now leads to two generations of leptons 
and baryons, but these are fixed by the several Higgs expectation values which
are approximately determined by the minimisation of the semiclassical potential, allowing
for small corrections $\Delta$ due to quantum effects. The resulting masses give
hope that the more realistic scenario, based on five properties, may be able to 
account for the known masses and mixings of the three established generations
and also escape the experimental exclusion of the single Higgs boson
of the standard model. But that is a much more difficult problem which will require
the use of and automation by mathematical packages as well as detailed analysis.

\section{A purely leptonic scheme} %Section II
Before getting embroiled with a pair of $\zeta$, we refer the reader to the
appendix, where the case of only one complex $\zeta$ is treated. There is a 
simplifying lesson to be learnt there if one is concerned with left and right 
field components and their gauging. The result of that analysis, so far as the
interaction with the Higgs superfield is concerned, is that we can, without
error, take the fermion source superfield $\Psi$ to be a Dirac field (left +
right) and a function of $\bar{\zeta}$ multiplied by powers of $\bar{\zeta}\zeta$, 
ensuring that it obeys some self-duality constraint. The (self-dual) Higgs 
superfield $\Phi$ is to be handled in the normal way and will therefore
couple the left and right fields together.

\subsection{Expansion in two properties}
Thus our starting point is the pair of properties, $\zeta^0$ signifying 
neutrinicity and $\zeta^4$ signifying charge -1 leptonicity, and a complete
disregard of strong interactions. (The full Sp(10) theory includes strong
interactions via three
further properties $\zeta^{1,2,3}$ signifying colour down-type quarks.)
The anti-selfdual fermions' superfield and adjoint, properly
normalized and simplified, consists of a single generation,
\begin{eqnarray}
 \bar{\Psi} &=& [\bar{\zeta}_0(1-\bar{\zeta}_4\zeta^4)\nu +
               \bar{\zeta}_4(1-\bar{\zeta}_0\zeta^0)\ell]/\sqrt{2},\nonumber \\
 \overline{\Psi} &\equiv& [\bar{\nu}(1-\bar{\zeta}_4\zeta^4)\zeta^0 +
               \bar{\ell}(1-\bar{\zeta}_0\zeta^0)\zeta^4]/\sqrt{2}
\end{eqnarray}
while the anti-selfdual Higgs superfield,
\begin{equation}
 \Phi = Y(1-\bar{\zeta}_0\zeta^0\bar{\zeta}_4\zeta^4)/\sqrt{2} +
        Z^0(\bar{\zeta}_0\zeta^0-\bar{\zeta}_4\zeta^4)/\sqrt{2} + 
        Z^+(\bar{\zeta}_0\zeta^4) + Z^-(\bar{\zeta}_4\zeta^0),
\end{equation}
contains a weak isosinglet ($Y$) and isotriplet ($Z^+,Z^0,Z^-$).
In that respect it is badly in conflict with the standard model. However bear 
in mind that the inclusion of the strong chromicity triplet fixes this problem
and cures the lack of insufficient generations. The purpose of the
present section is not to describe a totally realistic scenario but to 
comprehend how the masses of fermions and bosons are linked to the 
parameters arising from a renormalizable superHiggs potential comprising 
{\em two} sets of Higgs component fields $Y$ and $Z$ {\em interacting with 
one another} and with the fermions by means of a {\em single} Yukawa coupling
constant $h$. [In the following section we do include some semblance of 
strong interactions, involving several Higgs component fields and thereby cure 
{\em some} of the deficiencies of this section.]

The fermionic Lagrangian arises from the usual combination
\[
\hspace{-2cm}
 {\cal L}_\Psi=\int\,d^2\bar{\zeta}d^2\zeta \,\overline{\Psi}(i\gamma.\partial
       -\sqrt{2}h \Phi]\Psi
\]
\begin{equation}\hspace{-1.2cm}
  = \bar{\nu}[i\gamma.\partial\!-\!h(Y\!\!+\!Z^0/2)]\nu +
      \bar{\ell}[i\gamma.\partial\!-\!h(Y\!\!-\!Z^0/2)]\ell -
     h(Z^+\bar{\nu}\ell\!+\!Z^-\bar{\ell}\nu)/\sqrt{2}
\end{equation}
while the renormalizable scalar potential consists of up-to-quartic terms
\footnote{The negative sign in front is an artifact of the fermionic integration 
and is really of no fundamental consequence; the correct signs appear in eq (4)}:
\begin{eqnarray}
 {\cal L}_\Phi &=&-\int d^2\bar{\zeta}d^2\zeta\,[(\partial\Phi)^2/2
   +\mu^2\Phi^2/2 +\sqrt{2}f\Phi^3/3+g\Phi^4/6]\nonumber\\
 &=& [(\partial Y)^2+(\partial Z^0)^2]/2+\partial Z^+.\partial Z^-
    +\mu^2[Y^2/2 +Z^{02}/2+Z^+Z^-]\nonumber\\
& & -f[Y^3/2 +YZ^{02}\!+2YZ^+Z^-] - g[Y^4/6 +Y^2Z^{02}/2 +\!Y^2Z^+Z^-].
\end{eqnarray}

\subsection{Expectation values}
There are five equations of motion: for the one charged and two uncharged Higgs fields
and the two leptons. Minimising the scalar potential in (4) so that the two uncharged
fields $Y$ and $Z^0$ acquire expectation values,
\[
Y=\langle Y\rangle + {\cal Y}\equiv y +{\cal Y},\quad{\rm and}\quad
Z^0=\langle Z^0\rangle + {\cal Z}\equiv z +{\cal Z},
\]
we arrive at two constraints. With just the classical quartic term, as in (4)
the following conditions have to hold exactly:
\begin{equation}
 f(3y^2/2+z^2)+g(2y^3/3+yz^2)-\mu^2y =  0,\quad 
 2fyz + gy^2z-\mu^2z = 0.
\end{equation}
At the same time the mass$^2$ matrix for the $\cal Y$ and $\cal Z$ fields reads
\begin{equation}
   \left( \begin{array}{cc}
         -\mu^2+3fy+2gy^2+gz^2 & 2z(f+gy) \\
         2z(f+gy) & -\mu^2+2fy+gy^2
         \end{array} \right)
\end{equation}
and we need to ensure the absence of tachyons. If we strictly try to solve eqs. (5) first
there are two possibilities: (a) that $z=0$ and $y\neq 0$ with $\mu^2=3fy/2+2gy^2/3$, 
whereupon $m_{\cal Y}^2 = 3fy/2 + 4gy^2/3,\quad m_{\cal Z}^2 = fy/2+gy^3/3$ can
both be positive -- so no tachyons -- or (b) $z\ne 0$ and $2fy+gy^2=\mu^2$, which
case unfortunately leads to tachyonic ${\cal Y}$ and $\cal Z$ fields and is therefore
unacceptable physically. Case (a) is perfectly fine physically but leads to degenerate
neutrino and lepton fields, so is uninteresting.

\subsection{Masses}
The conclusions above signify that really we should not limit ourselves to a classical 
quartic potential. Indeed we know that the the effective potential will include higher 
order terms in the participating fields from quantum loops; thus we shall relax the 
over-constrained system equations (5) and assume that they are only approximately true.
This will allow us to escape the straight-jacket of degenerate fermion masses which 
otherwise ensue for we need  two distinct nonzero values for $y$ and $z$ in (3).
We will therefore continue to assume that $z\neq 0$ and introduce a small parameter 
$\Delta$ having dimension of mass$^2$ into the second equation (5) that is simply meant
to encapsulate the result of higher order quantum corrections: $ \mu^2= 2fy+gy^2+\Delta$.
This proves sufficient to lead to a positive value for the determinant of the (${\cal Y},{\cal Z}$)
mass$^2$ matrix, namely, $\Delta(\Delta + fy+gy^2+\Delta)-4z^2(f+gy)^2$; further it
suggests that $z$ is of order $\sqrt{\Delta}$. No longer are the charged lepton and
neutrino mass degenerate. and all the results make physical sense. We shall adopt a similar strategy in the next section.

\section{Expansion in three properties}
There is much more substance in this section. To the pair $\zeta^{0,4}$ of leptonic properties 
which encapsulate weak isospin we shall attach a third (charged coordinate) $\zeta^5$ 
which mimics strong interactions but without colour and which can be likened to
the colourless product of (anti)red/blue/green down-quarks;  it carries fermion
number $F=-1$  and charge $Q=1$ \footnote{In previous papers we reserved the labels 
$\zeta^{1,2,3}$ for colour and shall continue to do so here, thereby ignoring them.}.  
The table below summarises the quantum number attributes:
\begin{table}[h]
\begin{center}
 \begin{tabular}{|c|c|c|c|c|}  \hline
     Property&$Q$ & $T_3$ & $L$ & $F$ \\
 \hline
 $\zeta^0$ & 0 & 1/2 & 1 & 1 \\
 $\zeta^4$ & -1 & -1/2 & 1 & 1\\
 $\zeta^5$ & 1 & 0 & 0 & -1\\
 \hline
 \end{tabular}
 \end{center}
 \caption{Charge $Q$, weak isospin $T_3$, lepton number $L$ and fermion number $F$
 of the three fundamental properties.}
\end{table}

Since a superfield is to be expanded in powers of the three $\zeta$ and three 
$\bar{\zeta}$ we can come across combinations like
$\bar{\zeta}_0\bar{\zeta}_4\zeta^5$ having $F=3, Q=-2$ or 
$\bar{\zeta}_0\bar{\zeta}_5\zeta^4$ having $F=-1,Q=2$; these are unpleasant
states which have never been observed experimentally, so we need to find
a way of excising them. This can be done by requiring the superfield to be
anti-selfdual, associated with reflection about the cross-diagonal as explained in
reference \cite{DJW91}. For instance the action of the duality operation (denoted by $^\times$)
on these combinations is
\[ 
   (\bar{\zeta}_0\bar{\zeta}_4\zeta^5)^\times = \bar{\zeta}_0\bar{\zeta}_4\zeta^5,
   \qquad (\bar{\zeta}_0\bar{\zeta}_5\zeta^4)^\times = \bar{\zeta}_0\bar{\zeta}_5\zeta^4.
\]
This does not remove all exotic states, for instance the leptobaryon combinations,
\[
\zeta^0\zeta^4\, {\rm with~}Q=-1,\quad \bar{\zeta}_4\zeta^5\,{\rm with~} Q=2,
\]
survive, etc.
Upon imposing antiduality on both the superfields, $\Phi$ for bosons and $\Psi$
for fermions, we arrive at expansions that comprise {\em two} generations of fermions (each
consisting of a neutrino, charged lepton and `proton') and {\em five} uncharged Higgs fields (three
weak isosinglets $A,B,U$, one isotriplet  $C^0$ and one isodoublet $D^0$) plus a few charged
partners. Properly normalized, the antidual expansions read in full:
\begin{eqnarray}
\sqrt{2}\Psi & = &\quad \bar{\zeta}_0\nu(1+\bar{\zeta}_4\zeta^4\bar{\zeta}_5\zeta^5) 
                                       + \bar{\zeta}_0\nu' (\bar{\zeta}_4\zeta^4+\bar{\zeta}_5\zeta^5)
           \nonumber \\
           & & +  \bar{\zeta}_4\ell(1+\bar{\zeta}_0\zeta^0\bar{\zeta}_5\zeta^5)
                  + \bar{\zeta}_4\ell ' (\bar{\zeta}_0\zeta^0+\bar{\zeta}_5\zeta^5)
            \nonumber \\
           & & +  p^c \zeta^5(1+\bar{\zeta}_0\zeta^0\bar{\zeta}_4\zeta^4)
                  + {p^c}' \zeta^5(\bar{\zeta}_0\zeta^0+\bar{\zeta}_4\zeta^4) ,        
\end{eqnarray}
\begin{eqnarray}
\sqrt{2}\bar{\Phi}&  = & \quad A(1-\bar{\zeta}_0\zeta^0\bar{\zeta}_4\zeta^4\bar{\zeta}_5\zeta^5)
                                       +U(\bar{\zeta}_5\zeta^5-\bar{\zeta}_4\zeta^4\bar{\zeta}_0\zeta^0)
            \nonumber \\ 
           & & +  B(\bar{\zeta}_0\zeta^0+\bar{\zeta}_4\zeta^4)(1-\bar{\zeta}_5\zeta^5)/\sqrt{2} 
           \nonumber \\
           & & +  (C^+\bar{\zeta_0}\zeta^4+C^-\bar{\zeta}_4\zeta^0
                       +C^0(\bar{\zeta}_0\zeta^0-\bar{\zeta}_4\zeta^4)/\sqrt{2})(1+\bar{\zeta}_5\zeta^5)
          \nonumber \\
          & & + (D^0\zeta^4\zeta^5\!-\!\bar{D}^0\bar{\zeta}_5\bar{\zeta}_4)(1\!+\!\bar{\zeta}_0\zeta^0)
             + (D^-\zeta^0\zeta^5\!+\!D^+\bar{\zeta}_0\bar{\zeta}_5)(1\!+\!\bar{\zeta}_4\zeta^4)
          \nonumber \\
          & & + (S^+\zeta^0\zeta^4 + S^-\bar{\zeta}_0\bar{\zeta}_4)(1+\bar{\zeta}_5\zeta^5) 
          + (T^{--}\bar{\zeta}_4\zeta^5+T^{++}\bar{\zeta}_5\zeta^4)(1+\bar{\zeta}_0\zeta^0)
          \nonumber \\
          & & + (T^{-}\bar{\zeta}_0\zeta^5+T^{+}\bar{\zeta}_5\zeta^0)(1+\bar{\zeta}_4\zeta^4)
\end{eqnarray}
Thus we encounter three uncharged isosinglet Higgs $A,B,U$ plus one charged combination
$S^+$; two isospin doublets ($D^+, D^0$), ($T^{++},T^+$) and one isotriplet $(C^+, C^-, C^0)$.
We can regard $D$ and $T$ as leptoquark fields; note that only the expectation value of the
uncharged $D^0$ has a bearing on lepton-baryon mass splitting.

Expanding the various parts of the renormalizable Lagrangian for the superHiggs field 
purely on its own, viz.
\begin{equation}
{\cal L}= \int d^3\bar{\zeta}\,d^3\zeta\,\,[(\partial\Phi)^2/2+\mu^2\Phi^2/2-\sqrt{2}f\Phi^3/3-g\Phi^4/4],
\end{equation}
we encounter the following complicated combinations of terms::
\begin{eqnarray}
\int d^3\bar{\zeta}\,d^3\zeta\,\,\Phi^2&=&A^2+B^2+U^2+{C^0}^2+2C^+C^-+\nonumber \\
                &&2(D^+D^-\!+\!\bar{D}^0D^0\!+\!S^+S^-\!+\!T^+T^-\!+\!T^{++}T^{--});
\end{eqnarray}
\begin{eqnarray}
\sqrt{2}\int d^3\bar{\zeta}\,d^3\zeta\,\,\Phi^3/3&=&A[A^2/2+U^2+B^2+{C^0}^2+2C^+C^-+\nonumber\\
           &&\hspace{-1cm}2\bar{D}^0D^0+2D^+D^-+2S^+S^-+2T^+T^-+2T^{++}T^{--}]+\nonumber\\
           &&\hspace{-1cm} U(C^+C^-+{C^0}^2/2-B^2/2+S^+S^-)+\nonumber \\
           &&\hspace{-1cm}B(\bar{D}^0D^0+D^+D^-+T^{++}T^{--}+T^+T^-)/\sqrt{2}+\nonumber\\
           &&\hspace{-1cm}C^0(\bar{D}^0D^0-D^+D^-+T^{++}T^{--}-T^+T^-)/\sqrt{2}+\nonumber\\
           &&\hspace{-1cm}S^+(\bar{D}^0T^--D^+T^{--})+S^-(D^0T^+-D^-T^{++})-\nonumber\\
           &&\hspace{-1cm}C^+(\bar{D}^0D^-+T^+T^{--})-C^-(D^0T^++T^-T^{++});
\end{eqnarray}
\begin{eqnarray}
\int d^3\bar{\zeta}\,d^3\zeta\,\,\Phi^4/6&=&A^2[A^2/6+(B^2+U^2+C^{o2})/2+C^+C^-+\nonumber\\
        &&\hspace{-1cm}\bar{D}^0D^0+D^+D^-+S^+S^-+T^+T^-+T^{++}T^{--}]+\nonumber\\
        &&\hspace{-1cm} AU(C^+C^-+{C^0}^2/2-B^2/2+S^+S^-)+\nonumber \\
        &&\hspace{-1cm}AB(\bar{D}^0D^0+D^+D^-+T^{++}T^{--}+T^+T^-)/\sqrt{2}+\nonumber\\	         
        &&\hspace{-1cm}A{C^0}(\bar{D}^0D^0-D^+D^-+T^{++}T^{--}-T^+T^-)/\sqrt{2}+\nonumber\\
        &&\hspace{-1cm}AS^+(\bar{D}^0T^--D^+T^{--})+AS^-(D^0T^+-D^-T^{++})-\nonumber\\
       &&\hspace{-1cm}AC^+(\bar{D}^0D^-+T^+T^{--})-AC^-(D^0T^++T^-T^{++}).
\end{eqnarray}

As we are looking for spontaneously broken solutions which are nevertheless charge-conserving,
the only expectation values we can allow belong to the fields $A,B,U,C^0,D^0$. We therefore
make the expansions
$$
A=\langle A\rangle +{\cal A} \equiv  {a} + {\cal A},\quad B=\langle B\rangle+{\cal B} \equiv  b + {\cal B},
\quad U=\langle U\rangle +{\cal U} \equiv  u + {\cal U},
$$
$$
C^0=\langle C^0\rangle +{\cal C} \equiv  c + {\cal C},\quad 
 D^0=\langle D^0\rangle+{\cal D} \equiv  d/\sqrt{2} + {\cal D},
$$
and ascertain the minimum of the quartic potential; this occurs where
\begin{eqnarray}
\mu^2a &=&f[3a^2/2+b^2+u^2+c^2+d^2]+\nonumber\\
            & & g[a(2a^2/3\!+\!b^2\!+\!c^2\!+\!d^2\!+\!u^2)+u(c^2-b^2)/2+d^2(b+c)/2\sqrt{2}],
\end{eqnarray}
\begin{equation}
\mu^2b =f[2ab-ub+d^2/2\sqrt{2}]+g[ab(a-u)+ad^2/2\sqrt{2}],
\end{equation}
\begin{equation}
\mu^2u =f[2au+(c^2-b^2)/2]+g[a^2u+a(c^2-b^2)/2],
\end{equation}
\begin{equation}
\mu^2c =f[2ac+uc+d^2/2\sqrt{2}]+g[ac(a+u)+ad^2/2\sqrt{2}],
\end{equation}
\begin{equation}
\mu^2d =f[2ad+d(b+c)/\sqrt{2}]+g[a^2d+ad(b+c)/\sqrt{2}].
\end{equation}
The following combinations are obiquitous so we shall abbreviate them for future use:
$$F\equiv f+ga,\,G\equiv 2fa+ga^2.$$

Firstly let us study the fermion mass matrix hailing from the interaction
$\int d^3\bar{\zeta} d^3\zeta\,\bar{\Psi}[m+\sqrt{2}h\langle \Phi \rangle]\Psi $. 
In the space of states $(\nu,\nu',\ell,\ell',p^c,{p^c}')$, we obtain the elements:
\begin{equation}
       h\left( \begin{array}{cccccc}
        \alpha-\frac{b+c}{2\sqrt{2}}&\frac{u}{2}+\frac{b-c}{2\sqrt{2}}&0&0&0&0\nonumber\\
        \frac{u}{2}+\frac{b-c}{2\sqrt{2}}&\alpha&0&0&0&0\nonumber\\
        0&0&\alpha-\frac{b-c}{2\sqrt{2}}&\frac{u}{2}+\frac{b+c}{2\sqrt{2}}&-\frac{d}{2\sqrt{2}}&
        -\frac{d}{2\sqrt{2}}\nonumber\\
        0&0&\frac{u}{2}+\frac{b+c}{2\sqrt{2}}&\alpha&-\frac{d}{2\sqrt{2}}&0\\
       0&0&-\frac{d}{2\sqrt{2}}&-\frac{d}{2\sqrt{2}}&\alpha-\frac{u}{2}&\frac{b}{\sqrt{2}}\nonumber\\
        0&0&-\frac{d}{2\sqrt{2}}&0&\frac{b}{\sqrt{2}}&\alpha
         \end{array} \right),
\end{equation}
where $\alpha=a+m/h$. Thus it is essential to look for solutions where $d\neq 0$ 
in order to separate the lepton and baryon sectors, and $u+b-c\neq 0$ if 
we wish to separate generations.

Next let us examine the mass matrices of the various scalar mesons, which must 
conform {\em approximately} to the above restrictions on the expectation values since
we know that the effective potential will disturb conditions (13) to (17). It is important 
for us to ensure there are no tachyonic particles. Start with the Higgs multiplets 
contained in $\Phi$, of which the only unmixed state is $T^{++}$. It has
\begin{equation}
M^2_{T^{++}} = G+(b+c)F/\sqrt{2}-\mu^2.
\end{equation}
On the other hand the states $(C^+,-D^+)$ and separately $(S^+,T^+)$ mix with
identical mass$^2$ matrices and therefore form degenerate pairs:
\begin{equation}
  \left( \begin{array}{cc}
         G+uF-\mu^2 & dF/\sqrt{2} \\
         dF/\sqrt{2} & G+(b-c)F\sqrt{2}-\mu^2
         \end{array} \right).
\end{equation}
Next, the full set of(symmetric) 5$\times$5 matrix elements for the uncharged 
 quantum states $(A,B,C^0,U,(D^0+\bar{D}^0)/\sqrt{2})$ read:
\begin{eqnarray}
 M^2_{AA} &=& aF+G+g(b^2+c^2+u^2+d^2)-\mu^2,\\
 M^2_{AB} &=& M^2_{BA}=2bF-gbu+gd^2/2\sqrt{2}],\\
 M^2_{AC}&=&M^2_{CA}=2cF+gcu+gd^2/2\sqrt{2}],\\
 M^2_{AU}&=&M^2_{UA}=2uF+g(c^2-b^2)/2],\\
 M^2_{AD}&=&M^2_{DA}=2dF+gd(b+c)/\sqrt{2}],\\
 M^2_{BB}&=&G-uF-\mu^2,\quad M^2_{BC}=M^2_{CB}=0,\\
 M^2_{BU}&=&M^2_{UB}=-bF,\quad M^2_{BD}=M^2_{DB}=dF/\sqrt{2},\\
 M^2_{CC}&=&G+uF-\mu^2,\\
 M^2_{CU}&=&M^2_{UC}=cF,\quad M^2_{CD}=M^2_{DC}=dF/\sqrt{2},\\
M^2_{UU}&=&G-\mu^2,\quad M^2_{UD}=M^2_{DU}=0,\\
M^2_{DD}&=&G+(b+c)F/\sqrt{2}-\mu^2.
\end{eqnarray}

The most practical strategy for analysing the remaining conditions which ensure
that we get physically acceptable solutions is as follows. There are two free parameters
in our scheme (apart from coupling constants and masses) that are {\em not} determined by
minimising the effective potential and these amount to fixing somehow two of the
five expectation values of the super-Higgs components -- which must therefore be
found by other considerations, such as dynamical symmetry breaking or the inclusion
of quantum corrections. To gain an idea first
of what is possible, consider the particular case where $b=c$ and $u=0$. One then
readily checks that no tachyon states arise provided the conditions below are met:
\begin{eqnarray}
  G-\mu^2&\geq\sqrt{2}|bF|\\
  G-\mu^2&\geq |dF|/\sqrt{2}.
\end{eqnarray}
and the eigenvalues of the $3\times 3$ mass$^2$ for the coupled fields 
$A,(B+C^0),(D^0+\bar{D}^0)$
$$
  \left( \!\!\!\begin{array}{ccc}
  aF+G+g(2b^2\!+\!d^2)-\mu^2 & 2bF+gd^2/2\sqrt{2} & 2d(F+gb/\sqrt{2})\\
   2bF+gd^2/2\sqrt{2} &G-\mu^2 & dF/\sqrt{2}\\
   2d(F+gb/\sqrt{2})&dF/\sqrt{2}& G+\sqrt{2}bF -\mu^2
   \end{array}\!\!\! \right)
$$
are positive; and this is indeed possible. Inclusion of additional corrections arising when $b\neq c$
and $u\neq 0$ does not disturb the conclusion that there are no tachyons. The last consideration,
which is the least important, is how well the minimisation conditions of the classical quartic
potential, eqs (13)-(17) are satisfied, since we are certain that they are liable to be muddied
by quantum loops \footnote{If the constraints (13) to(17) are strictly imposed we arrive at the
ridiculous result that there are no tachyons but all particles are massless}.
For purposes of illustration, we can show that `acceptable' masses can arise for some 
suitable inputs (masses in GeV for dimensionful quantities):
\[
{\rm Dimensionless~couplings:}\, g=2,\,h=1;
\quad{\rm Dimensionful~coupling:}\,f=0.3;
\]
\[
{\rm Masses:}\, m=0;\,\,-\mu^2=1.2;
\]
\[
{\rm Expectation~values:}\,a=0.5,\,b=c=0.5,\,u=-0.5,\,d=-0.3.
\]
The consequence of these rough choices is the following sets of particle masses
for fermions and Higgs mesons:
\[
{\rm Neutrinos:}\,m_\nu=0.02,\,m'_\nu=0.63;
\]
\[
{\rm Leptons/Baryons:}\,m_\ell=0.30,\,m'_\ell=0.44,\,m_p=0.53,\,m'_p=0.98;
\]
\[
{\rm Doubly~charged~Higgs~meson:}\,m^2_{T^{++}}=4.35;
\]
\[
{\rm Singly~charged~Higgs~mesons:}\,m^2_{C^+}=m^2_{S^+}=1.78,
       \,m^2_{D^+}=m^2_{T^+}=3.09;
\]
\[
{\rm Uncharged~Higgs:}\,m^2_A\!=\!9.00,\,m^2_D\!=\!3.01,\,m^2_C\!=\!4.35,
      \,m^2_U\!=\!2.34,\,m^2_B\!=\!0.09,
\]
which values can be scaled up and adjusted further if needed by varying the inputs.
The small deviations from the classical quartic potential constraints can thereby be 
estimated. For instance, typically $\Delta \sim u(G-\mu^2)\simeq 2.9$ enters eq (15)
and corresponds to the magnitude expected from quantum effects.

These results are not ridiculous for a very crude 2-generation model. They give hope
that a more realistic model based on 5 property coordinates might produce reasonable
values for the known generations of colourless particles. But in any case schemes of
this type have fewer parameters than the standard model; thus the masses and mixings
of fermions in conventional model contain at least 18 independent constants ignoring
CP violation, whereas 5-property schemes only contain at most 9 Higgs expectation 
values and only {\em one} Yukawa coupling. The latter are therefore worth exploring.
Also of great interest is to what extent the initial anti-selfduality symmetry of the
superfields is preserved by quantum corrections. This can be verified by looking
at the one-loop corrections to the self-energies of the participating fields and the
3-point Yukawa couplings to see if they are renormalized by exactly the same amount; 
but this is a major study and must be left to a future investigation.

\section*{Appendix A - Simplifying treatment of the left/right fields}
Let us focus on a single complex $\zeta$, corresponding to an Sp(2) group
structure, associated with a single fermion number. 
Strictly, the fermion superfield $\Psi(x,\zeta,\bar{\zeta})$ must depend 
on property and antiproperty and also involve both left and right
chirality components. Recalling the usual charge conjugation relation,
$\psi^c = C \tilde{\bar{\psi}}$, we note the basic result, ${\psi_L}^c
= {\psi^c}_R$. It follows that if the source superfield is considered
to be totally left-handed,
\[
\Psi(x,\zeta,\bar{\zeta}) = \bar{\zeta}\psi_L + {\psi^c}_L\zeta,
\]
it does in reality incorporate both chirality components. The superfield
adjoint then has to be defined as
\[
\overline{\Psi}(x,\zeta,\bar{\zeta}) = \overline{{\psi}_R}\zeta -\bar{\zeta}\,
\overline{{\psi^c}_R},
\]
in order to ensure
\[
\overline{\Psi}\Psi = \zeta\bar{\zeta}\,[\overline{\psi_R}\psi_L +
 \overline{{\psi^c}_R}{\psi^c}_L] 
= \zeta\bar{\zeta}\,[\overline{\psi_R}\psi_L + \overline{\psi_L}\psi_R],
\]
in tandem with the Grassmannian integration rule:
$\int d\bar{\zeta}\,d\zeta \,\, \zeta\bar{\zeta} = 1$.

Bearing in mind that $\Psi$ is overall {\em Bose}, the kinetic energy
term needs to be taken as the bilinear combination,
\begin{flushleft}
$\Psi_\alpha (iC^{-1}\gamma.\partial)^{\alpha\beta}
 (\zeta\partial_\zeta - \bar{\zeta}\partial_{\bar{\zeta}})\Psi_\beta = 
 \zeta\bar{\zeta}\,[\overline{{\psi}_L}i\gamma.\partial\psi_L + 
   \overline{{\psi^c}_L}i\gamma.\partial{\psi^c}_L] =
 \zeta\bar{\zeta}\,[\overline{{\psi}_L}i\gamma.\partial\psi_L + 
   \overline{\psi_R}i\gamma.\partial\psi_R]$.
\end{flushleft} 
These manoeuvres are of great importance if one is interested in
gauging one or other of the chiralities. For instance a local 
left-field phasing would engender the gauge combination 
$A\bar{\zeta}\partial_{\bar{\zeta}}$ {\em only}, while the right-field
is accompanied by $A\zeta\partial_\zeta$, etc.  Turning to the
superfield scalar $\Phi$, which couples left with right, it is
simply described by the the self-dual combination
\[
\Phi(x,\zeta,\bar{\zeta}) = [1-\bar{\zeta}\zeta]H /\sqrt{2}
\]
so the Yukawa interaction with the source superfield is nothing but
the usual interaction:
\begin{flushleft}
 $\sqrt{2}h\int d\bar{\zeta}\,d\zeta\,\bar{\Psi}\Phi\Psi = 
 h\int d\bar{\zeta}\,d\zeta\,(\zeta\bar{\zeta})\,
 [\overline{\psi_R}\psi_L + \overline{\psi_L}\psi_R](1-\zeta\bar{\zeta})
 H = h\bar{\psi}\psi H/\sqrt{2}.$
\end{flushleft}
Upon including the kinetic term of the source and adding that of
the scalar field, namely
\[
\int d\bar{\zeta}\,d\zeta\,(\partial\Phi)^2/2 = (\partial H)^2/2,
\]
plus the renormalizable self-interaction
\[
 \int d\bar{\zeta}\,d\zeta\,[\mu^2\Phi^2/2-\lambda\Phi^4/4] =
 \mu^2H^2/2 - \lambda H^4/4,
\]
we are back to the original Higgs model.  

This may seem like a sledgehammer to crack a nut (and it probably is)
but, when we add further properties, exciting new possibilities arise.
As we are only interested in the mass generation mechanism through 
coupling to the scalar superfield and are not concerned about the
gauge field, we can bypass the left-right fuss by combining the
components into the Dirac superfields $\Psi = \bar{\zeta}\psi,
\overline{\Psi} = \bar{\psi}\zeta$ and just leave it at that, apart from 
products with polynomials of $\bar{\zeta}\zeta$ which can occur with more 
than one property when one is imposing self-duality on the superfields.
That is the strategy we have adopted in sections 2 and 3.

\section*{Acknowledgements}

I would like to express my thanks to Peter Jarvis and Paul Stack for numerous
discussions and assistance.

\end{document}